\documentclass[twocolumn,prl,superscriptaddress,longbibliography]{revtex4-1}
\usepackage{graphicx,amsmath,amssymb}
\usepackage[colorlinks=true, citecolor=blue, urlcolor=blue, linkcolor=red]{hyperref}
\renewcommand{\section}[1]{{\par\it #1.---}\ignorespaces}

\begin{document}
\title{Non-Markovian Sensing of a Quantum Reservoir}
\author{Wei Wu}
\affiliation{School of Physical Science and Technology, Lanzhou University, Lanzhou 730000, China}
\author{Si-Yuan Bai}
\affiliation{School of Physical Science and Technology, Lanzhou University, Lanzhou 730000, China}
\author{Jun-Hong An}
\email{anjhong@lzu.edu.cn}
\affiliation{School of Physical Science and Technology, Lanzhou University, Lanzhou 730000, China}

\begin{abstract}
Quantum sensing explores protocols using the quantum resource of sensors to achieve highly sensitive measurement of physical quantities. The conventional schemes generally use unitary dynamics to encode quantities into sensor states. In order to measure the spectral density of a quantum reservoir, which plays a vital role in controlling the reservoir-caused decoherence to microscopic systems, we propose a nonunitary-encoding optical sensing scheme. Although the nonunitary dynamics for encoding in turn degrades the quantum resource, we surprisingly find a mechanism to make the encoding time a resource to improve the precision and to make the squeezing of the sensor a resource to surpass the shot-noise limit. Our result shows that it is due to the formation of a sensor-reservoir bound state. Enriching the family of quantum sensing, our scheme gives an efficient way to measure the quantum reservoir and might supply an insightful support to decoherence control.
\end{abstract}

\maketitle
%%%%%%%%%%%%%%%%%%%%%%%%%%%%%%%%%%%%%%%%%%%%%%%%%%%%%
\section{Introduction}
%%%%%%%%%%%%%%%%%%%%%%%%%%%%%%%%%%%%%%%%%%%%%%%%%%%%%
Quantum sensing pursues highly precise measuring of physical quantities by using quantum effects of sensors in specific encoding and measurement protocols~\cite{RevModPhys.89.035002,RevModPhys.90.035005,doi:10.1116/1.5119961,Lachance-Quirion425}. The precision of any measurement obeying classical physics is bounded by the shot-noise limit (SNL). Quantum sensing allows us to attain a precision that surpasses the SNL by using quantum resources, such as entanglement~\cite{PhysRevLett.122.123605,PhysRevLett.121.160502,Unternahrer:18,Zou6381,Nagata726,Daryanoosh2018} and squeezing~\cite{PhysRevD.23.1693,PhysRevLett.118.140401,PhysRevLett.119.193601}. It has inspired many fascinating applications in gravitational wave detection \cite{PhysRevLett.123.231107, PhysRevLett.123.231108}, quantum radar \cite{PhysRevLett.123.203601,Gregoryeaay2652}, ultimate clocks \cite{YeJ2014,PhysRevLett.117.143004,Hosten}, quantum magnetometry \cite{PhysRevLett.113.103004,11} and optical lithography \cite{PAVEL2013259}.
These sensing schemes commonly resort the unitary dynamics of the sensors to encode the quantities into the sensor states. It is applicable only to measure classical systems \cite{PhysRevX.9.041023,PhysRevLett.122.090503,PhysRevLett.124.020501,PhysRevLett.124.080503,Wang2019,PhysRevResearch.1.032024,McCormick2019,Chabuda2020,PhysRevApplied.13.024037,Haase_2018,PhysRevLett.123.040402,PhysRevLett.124.030501,PhysRevLett.124.010507,PhysRevA.99.033807,PhysRevA.100.032318}. To sense a quantum system, the quantized sensor-system coupling for encoding the quantities definitely makes the dynamics of the sensor nonunitary. A natural question is: how can one generalize the well-developed unitary-encoding quantum sensing schemes to the nonunitary case?

Recently, there is an increasing motivation to develop the nonunitary quantum sensing scheme, especially in the setting of measuring a quantum reservoir. The inevitable interactions with the reservoir would cause any microscopic system to lose its quantum coherence, which is called decoherence and is a main bottleneck to realize quantum computers and other quantum tasks \cite{Rivas_2014,RevModPhys.88.021002,RevModPhys.89.015001,LI20181}. Characterizing the system-reservoir coupling strength in different reservoir frequencies, the spectral density completely determines the decoherence. The full grasping to the features of the spectral density is a prerequisite for understanding \cite{PhysRevLett.118.100401} and controlling \cite{PhysRevLett.87.270405} the decoherence. However, in many situations, the spectral density cannot be microscopically derived from first principles \cite{RevModPhys.59.1,Yang_2016}. Thus, a measuring of the spectral density is strongly necessary. Inspired by the high-precision nature of quantum sensing, several schemes have been proposed to measure the spectral density~\cite{PhysRevLett.118.100401,PhysRevLett.123.230801,PhysRevB.101.104306,PhysRevA.97.012126,PhysRevA.97.012125,tamascelli2020quantum,SALARISEHDARAN2019126006,PhysRevA.101.032112}. A substantial difference of these schemes from the previous ones \cite{PhysRevX.9.041023,PhysRevLett.122.090503,PhysRevLett.124.020501,PhysRevLett.124.080503,Wang2019,PhysRevResearch.1.032024,McCormick2019,Chabuda2020,PhysRevApplied.13.024037,Haase_2018,PhysRevLett.123.040402,PhysRevLett.124.030501,PhysRevLett.124.010507,PhysRevA.99.033807,PhysRevA.100.032318} resides in that the quantities are encoded into the sensor state via nonunitary dynamics, which in turn degrades the quantum resources of the sensors. This is expected to deteriorate the performance of the sensing schemes with increasing the encoding time. We see in Ref.~\cite{PhysRevA.97.012126,PhysRevA.97.012125,tamascelli2020quantum,SALARISEHDARAN2019126006,PhysRevA.101.032112} that the sensing precision to the spectral density not only does not surpass the SNL, but also gets worse and worse with increasing the encoding time. Thus, how to surpass the SNL in the long encoding time condition is still an open question in sensing a quantum reservoir.

We here propose a nonunitary sensing scheme using a quantized single-mode optical field as a sensor to estimate the spectral density of a quantum reservoir. A mechanism to make the encoding time and the optical squeezing resources to surpass the SNL at any long encoding time is present based on the non-Markovian description of the nonunitary encoding dynamics. Our analysis reveals that the proposed mechanism works efficiently due to the formation of a bound state in the energy spectrum of the composite system consisting of the optical sensor and the reservoir. With this mechanism, our scheme eliminates the outstanding error-divergence problem of sensing the quantum reservoir in the long encoding time regime.

\section{Quantum parameter estimation}
To estimate the quantity $\theta$ of a system, we first prepare a quantum sensor in the state $\varrho_\text{in}$ and couple it to the system to encode $\theta$ into the sensor state $\varrho_\theta$. Then we measure certain sensor observables and infer the value of $\theta$ from the result. The inevitable errors make us unable to estimate $\theta$ precisely. According to quantum parameter estimation theory \cite{Liu_2019,PhysRevLett.72.3439}, the ultimate precision of $\theta$ is constrained by the quantum Cram\'{e}r-Rao bound $\delta\theta\geq 1/\sqrt{\upsilon \mathcal{F}_{\theta}}$, where $\delta\theta$ is the standard error of the estimate, $\upsilon$ is the number of repeated measurements, and $\mathcal{F}_{\theta}=\text{Tr}(\hat{L}_\theta^2\varrho_\theta)$ with $\hat{L}_\theta$ defined by $\partial_\theta\varrho(\theta)=(\hat{L}_\theta\varrho_\theta+\varrho_\theta \hat{L}_\theta)/2$ is the quantum Fisher information (QFI) characterizing the most information of $\theta$ extractable from $\varrho_\theta$. We set $\upsilon=1$ due to the independence of $\mathcal{F}_{\theta}$ on $\upsilon$. How to maximize the QFI by resorting to the proper initial state $\varrho_\text{in}$ and sensor-system interaction is of importance in quantum sensing. If $\delta\theta$ is proportional to $\bar{n}^{-1/2}$ or $\mathcal{F}_{\theta}\propto\bar{n}$, with $\bar{n}$ being the number of the used resource, then the precision is called the SNL. The SNL can be beaten by using quantum protocols~\cite{PhysRevX.9.041023,PhysRevLett.122.090503,PhysRevLett.124.020501,PhysRevLett.124.080503,Wang2019,PhysRevResearch.1.032024,McCormick2019,Chabuda2020,PhysRevApplied.13.024037,Haase_2018,PhysRevLett.123.040402,PhysRevLett.124.030501,PhysRevLett.124.010507,PhysRevA.99.033807,PhysRevA.100.032318}.

%%%%%%%%%%%%%%%%%%%%%%%%%%%%%%%%%%%%%%%%%%%%%%%%%%%%%
\section{Quantum sensing to a dissipative reservoir}
%%%%%%%%%%%%%%%%%%%%%%%%%%%%%%%%%%%%%%%%%%%%%%%%%%%%%
We are interested in uncovering the quantum superiority in sensing a quantum reservoir with infinite degrees of freedom. We choose a single-mode optical field as the quantum sensor and utilize the following sensor-reservoir coupling to encode the reservoir quantities into the sensor ($\hbar=1$):
\begin{equation}\label{Hamlt}
\hat{H}=\omega_{0}\hat{a}^{\dagger}\hat{a}+\sum_{k}\omega_{k}\hat{b}_{k}^{\dagger}\hat{b}_{k}+\sum_{k}g_{k}\big{(}\hat{a}\hat{b}_{k}^{\dagger}+\hat{a}^{\dagger}\hat{b}_{k}\big{)},
\end{equation}
where $\hat{a}$ and $\hat{b}_{k}$ are the annihilation operators of the sensor with frequency $\omega_0$ and the $k$th reservoir mode with frequency $\omega_k$, respectively, and $g_{k}$ is the sensor-reservoir coupling strength. Depending on the properties of the reservoir, the coupling is characterized by the spectral density in the continuum limit of the frequency $J(\omega)\equiv\sum_{k}g_{k}^{2}\delta(\omega-\omega_{k})=g(\omega)^2D(\omega)$, where $D(\omega)=\int dk/d\omega_k\delta(\omega-\omega_k)d\omega$, called the density of state, depends on the dispersion relation of the reservoir.

Using the Feynman-Vernon's influence functional method to trace over the state of the reservoir~\cite{PhysRevA.76.042127,PhysRevLett.109.170402,PhysRevE.90.022122}, we derive the non-Markovian master equation of the sensor:
\begin{equation}\label{maseq}
\dot{\varrho}(t)=-i\Omega(t)\big{[}\hat{a}^{\dagger}\hat{a},\varrho(t)\big{]}+\Gamma(t)\check{\mathcal{L}}[\varrho(t)],
\end{equation}
where $\check{\mathcal{L}}[\cdot]=2\hat{a}\cdot\hat{a}^{\dagger}-\{\hat{a}^{\dagger}\hat{a},\cdot\}$, $\Omega(t)\equiv-\mathrm{Im}[{\dot u}(t)/u(t)]$ is the renormalized frequency, and $\Gamma(t)\equiv-\mathrm{Re}[\dot{u}(t)/u(t)]$ is a time-dependent dissipation coefficient. The reservoir is assumed in vacuum initially and $u(t)$ is determined by
\begin{equation}\label{intdiff}
\dot{u}(t)+i\omega_{0}u(t)+\int_{0}^{t} \nu(t-\tau)u(\tau)d\tau=0,
\end{equation}
with $u(0)=1$ and $\nu(x)=\int_{0}^{\infty}J(\omega)e^{-i\omega x}d\omega $. Keeping the same Lindblad form as the Markovian master equation, Eq. \eqref{maseq} collects the non-Markovian effect in the time-dependent coefficients. We see from Eq. \eqref{intdiff} that $J(\omega)$ has been successfully encoded into $\varrho(t)$ by Eq. \eqref{maseq}.

Different from the widely used unitary evolution \cite{PhysRevX.9.041023,PhysRevLett.122.090503,PhysRevLett.124.020501,PhysRevLett.124.080503,Wang2019,PhysRevResearch.1.032024,McCormick2019,Chabuda2020,PhysRevApplied.13.024037,Haase_2018,PhysRevLett.123.040402,PhysRevLett.124.030501,PhysRevLett.124.010507,PhysRevA.99.033807,PhysRevA.100.032318}, the quantity encoding governed by Eq. \eqref{maseq} is a nonunitary dynamics of the sensor. Although it would cause decoherence to the sensor, we still can estimate the quantities in $J(\omega)$ in higher precision than the classical SNL via properly utilizing the quantum characters of the sensor. We consider the initial state of the sensor as the squeezed state $\varrho(0)=\hat{\mathcal D}(\alpha)\hat{\mathcal S}(r)|0\rangle\langle 0|\hat{\mathcal S}^\dag(r)\hat{\mathcal D}^\dag(\alpha)$, where $\hat{\mathcal D}(\alpha)=\exp(\alpha\hat{a}^{\dagger}-\alpha^{*}\hat{a})$ and $\hat{\mathcal S}(r)=\exp[r(\hat{a}^{2}-\hat{a}^{\dag2})/2]$, with $|0\rangle$ being the vacuum state. The total photon number $\bar{n}=|\alpha|^{2}+\sinh^{2}r$, which contains the ratio $\beta\equiv\sinh^{2}r/\bar{n}$ from the squeezing of the optical sensor and is regarded as the quantum resource of the scheme. The ratio $\beta$ varies from $0$ for a coherent state to $1$ for a squeezed vacuum state. The Gaussianity of the initial state is preserved during the evolution governed by Eq. \eqref{maseq}. The Gaussian states are those whose characteristic function is of Gaussian form \cite{RevModPhys.77.513,_afr_nek_2015}
$\chi ({\pmb\gamma })\equiv\text{Tr}[\varrho \hat{\mathcal D}({\gamma})]=\exp [-{\frac{1}{4}}{\pmb\gamma }^{\dag }{\pmb \sigma} {\pmb\gamma}-i{\mathbf{d}}^{\dag }K{\pmb\gamma }]$, where ${\pmb \gamma}=(\gamma,\gamma^*)^T$, $K=\text{diag}(1,-1)$, and the elements of the displacement vector ${\mathbf{d}}$ and the covariant
matrix ${\pmb \sigma}$ are $d_{i} =\text{Tr}[\varrho \hat{A}_{i}]$ and $\sigma _{ij} =\text{Tr}[\varrho \{\Delta \hat{A}_{i},\Delta \hat{A}_{j}^{\dag}\}]$
with $\hat{\mathbf{A}}=(\hat{a},\hat{a}^\dag)^T$ and $\Delta \hat{A}_{i}=\hat{A}_{i}-d_{i}$. The QFI for $\theta$ in $\varrho$ reads~\cite{_afr_nek_2018,_afr_nek_2015,Gao2014}
\begin{equation}\label{QtFS}
\mathcal{F}_{\theta}=\frac{1}{2}[\text{vec}(\partial_\theta {\pmb \sigma})]^\dag\mathcal{M}^{-1}\text{vec}(\partial_\theta {\pmb \sigma})+2(\partial_{\theta}\mathbf{d})^{\dagger}\pmb{\sigma}^{-1}\partial_{\theta}\mathbf{d},
\end{equation}
where $\mathcal{M}={\pmb\sigma}^*\otimes{\pmb\sigma}-K\otimes K$, with $\pmb{\sigma}^{*}$ being the complex conjugate of $\pmb{\sigma}$.

Taking the Ohmic-family spectral density $J(\omega)=\eta \omega(\omega/\omega_c)^{s-1}e^{-\omega/\omega_c}$ as an example, we reveal the performance of our non-Markovian sensing scheme. The dimensionless constant $\eta$ characterizes the sensor-reservoir coupling strength, the cutoff frequency $\omega_c$ characterizes the correlation time scale of the reservoir, and the exponent $s$ relating to the spatial dimension classifies the reservoir into sub-Ohmic when $0<s<1$, Ohmic when $s=1$, and super-Ohmic when $s>1$ \cite{RevModPhys.59.1}. They are the parameters we estimate from $\varrho(t)$. Solving Eq. \eqref{maseq}, we obtain \cite{SupplementalMaterial}
\begin{eqnarray}
\mathbf{d}(t)&=&[\alpha u(t),\alpha^{*}u^{*}(t)]^{T},\label{displ} \\
\pmb{\sigma}(t)&=&\left[
                  \begin{array}{cc}
                    1+2|u(t)|^{2}\sinh^{2}r & -u(t)^{2}\sinh(2r) \\
                    -u^*(t)^{2}\sinh(2r) & 1+2|u(t)|^{2}\sinh^{2}r \\
                  \end{array}
                \right].\label{covrc}
\end{eqnarray}
Then the QFI of the parameters in $J(\omega)$ can be calculated.

When the sensor-reservoir coupling is weak and the correlation time scale of the reservoir is smaller than that of the sensor, we apply the Markovian approximation to Eq. \eqref{intdiff} and obtain $u_\text{MA}(t)= e^{-[\kappa+i(\omega_0+\Delta(\omega_0))]t}$, with $\kappa=\pi J(\omega_0)$ and $\Delta(\omega_0)=\mathcal{P}\int _0^\infty {J(\omega)\over \omega_0-\omega}d\omega$. It causes $\Gamma_\text{MA}(t)= \kappa$ and $\Omega_\text{MA}(t)=\omega_0+\Delta(\omega_0)$~\cite{PhysRevE.90.022122}, which are equal to the ones in the Markovian master equation. The Markovian approximate QFI in the large-$\bar{n}$ limit reads
\begin{equation}
\mathcal{F}^\text{MA}_\theta(t)=2(1-\beta)(\partial_\theta\kappa)^2[\coth(\kappa t)-1]\bar{n}t^2,\label{MAFI}
\end{equation}
where $\theta=\eta$, $\omega_c$, or $s$. We have neglected the constant $\Delta(\omega_0)$, which is generally renormalized into $\omega_0$ \cite{RevModPhys.59.1}. One can see from Eq. \eqref{MAFI} that $\lim_{t\rightarrow\infty}\mathcal{F}^\text{MA}_\theta(t)=0$, which indicates that no information on $\theta$ is extractable from $\varrho(\infty)$, and thus the scheme breaks down in the long-encoding time condition. This is physically understandable because the steady state of Eq. \eqref{maseq} under the Markovian approximation uniquely being the vacuum state $\varrho(\infty)=|0\rangle\langle0|$ does not carry any message on $J(\omega)$. However, after optimizing $\mathcal{F}^\text{MA}_\theta(t)$ to time, we have
\begin{equation}
\max\mathcal{F}^\text{MA}_\theta\simeq0.65(\partial_\theta \ln\kappa)^2\bar{n} \label{Markscal}
\end{equation}
when $t\simeq0.80\kappa^{-1}$. The maximum \eqref{Markscal} is achieved when $\beta=0$ for the input state is the coherent state. This implies that no quantum superiority is delivered by the squeezing. The scaling relation of $\max\mathcal{F}^\text{MA}_\theta$ to $\bar{n}$ equals exactly to the classical SNL.

In the non-Markovian dynamics, the analytical QFI is complicated. We leave it to the numerical calculation. However, via analyzing the long-time behavior of $u(t)$, we may obtain an asymptotic form of QFI. A Laplace transform can convert Eq.~\eqref{intdiff} into
$\tilde{u}(z)=[z+i\omega_0+\int_0^\infty{J(\omega)\over z+i\omega}d\omega]^{-1}$. The solution of $u(t)$ is obtained by the inverse Laplace
transform of $\tilde{u}(z)$, which can be done by finding its pole from
\begin{equation}
y(E)\equiv\omega_0-\int_0^\infty{J(\omega)\over\omega-E}d\omega =E,~(E=iz).\label{eigen}
\end{equation}
Note that the roots $E$ of Eq. \eqref{eigen} are just the eigenenergies of Eq. \eqref{Hamlt} in the single-excitation space. Specifically, expanding the eigenstate as $|\Psi\rangle=(x\hat{a}^{\dagger}+\sum_{k}y_{k}\hat{b}_{k}^{\dagger})|0,\{0_k\}\rangle$ and substituting it into $\hat{H}|\Psi\rangle=E|\Psi\rangle$, with $E$ being the eigenenergy, we have $(E-\omega_{0})x=\sum_{k}g_{k}y_{k}$ and $y_{k}=g_{k}x/(E-\omega_{k})$. They readily lead to Eq. \eqref{eigen}. It reveals that, although the spaces with any excitation numbers are involved, the dynamics is uniquely determined by the energy spectrum in the single-excitation space. Since $y(E)$ is a decreasing function in the regime $E < 0$, Eq. \eqref{eigen} has one isolated root $E_b$ in this regime provided $y(0) < 0$. While $y(E)$ is ill defined when $E>0$, Eq. \eqref{eigen} has infinite roots in this regime forming a continuous energy band. We call the eigenstate of the isolated eigenenergy $E_b$ the bound
state \cite{PhysRevLett.123.040402}. After the inverse Laplace transform, we obtain \cite{SupplementalMaterial}
\begin{equation}
u(t)=Ze^{-iE_b t}+\int_{0}^{\infty}\frac{J(E)e^{-iE t}dE}{[E-\omega_{0}-\Delta(E)]^{2}+[\pi J(E)]^2},\label{eq:eq13}
\end{equation}
where the first term with $Z=[1+\int_0^\infty{J(\omega)d\omega\over(E_b-\omega)^2}]^{-1}$ is from the bound state, and the second one is from the band energies. Oscillating with time in continuously changing frequencies, the integral tends to zero in the long-time condition due to out-of-phase interference. Thus, if the bound state is absent, then $\lim_{t\rightarrow\infty} u(t)= 0$ characterizes a complete decoherence, while if the bound state is formed, then $\lim_{t\rightarrow\infty} u(t)=Ze^{-iE_b t}$ implies a decoherence suppression. We can evaluate that the bound state for the Ohmic-family spectral density
is formed if $\omega_0\leq \eta\omega_c\underline{\Gamma}(s) $, where $\underline{\Gamma}(s)$ is the Euler's gamma function.

In the absence of the bound state, it is natural to expect that $\mathcal{F}_\theta(t)$ tends to zero because $u(t)$ approaches zero. Consistent with the Markovian result, the sensing scheme in this case also breaks down. In the presence of the bound state, we substitute the long-time $u(t)$ into Eq. \eqref{QtFS} and obtain \cite{SupplementalMaterial}
\begin{equation}\label{anaF}
\mathcal{F}_\theta(t)\simeq4Z^2\Theta(\beta,\bar{n})(\partial_\theta E_b)^2t^2
\end{equation}
with $\Theta(\beta,\bar{n})=\frac{\bar{n}(1-\beta)[1-2Z^{2}(\sqrt{\bar{n}\beta(1+\bar{n}\beta)}-\bar{n}\beta)]}{1+4\bar{n}\beta Z^{2}(1-Z^{2})}+\frac{2Z^{2}\bar{n}\beta(1+\bar{n}\beta)}{1+2\bar{n}\beta Z^{2}(1-Z^{2})}$.
In the case of $\beta=0$, we have $\Theta(0,\bar{n})=\bar{n}$. In the general case, $\Theta(\beta,\bar{n})\simeq{\bar{n}\beta\over 1-Z^2}$ can be derived in the large-$\beta\bar{n}$ limit. It is remarkable to find that, in contrast to the cases of Markovian approximation and without the bound state, $\mathcal{F}_\theta(t)$ in the non-Markovian dynamics increases with time in power law when the bound state is formed. The scaling of $\mathcal{F}_\theta(t)$ to time is the same as the ideal Ramsey-spectroscopy based quantum metrology, where the parameter encoding is via unitary dynamics \cite{PhysRevLett.79.3865,Gefen2019}. The result reveals that, thanks to the bound state, even the parameters are encoded via nonunitary dynamics with the decoherence presented, the sensing to the reservoir still performs as ideal as the unitary-encoding scheme.  Furthermore, although scaling with $\bar{n}$ in a manner similar to that of the SNL achieved by the coherent state $\beta=0$, QFI still has a dramatic prefactor improvement by the optimal squeezing $\beta=1$. Therefore, the bound state has made both the encoding time and the squeezing act as resources in our sensing scheme.

\begin{figure*}
\centering
\includegraphics[angle=0,width=.9\textwidth]{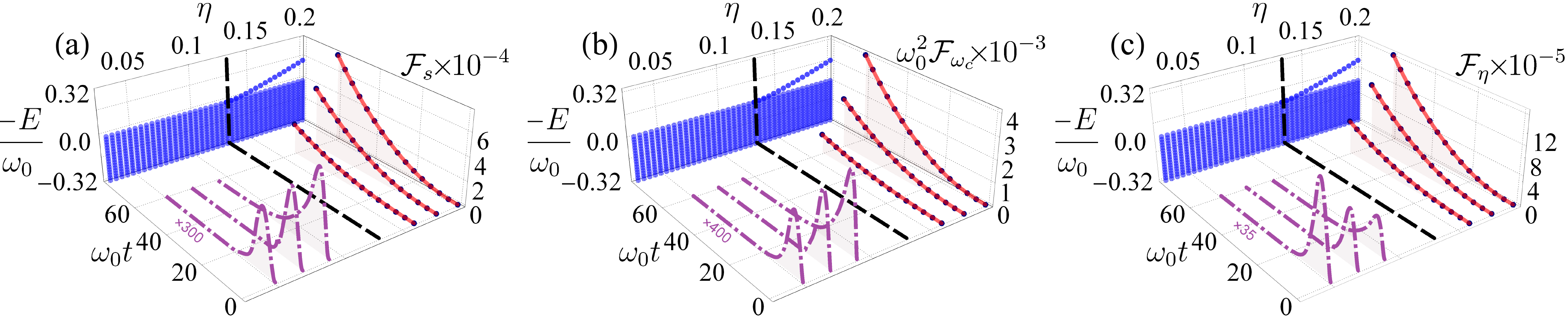}
\caption{ Evolution of $\mathcal{F}_{\theta}(t)$ in different $\eta$ obtained by numerically solving Eq.~\eqref{intdiff}. The purple dot-dashed lines are magnified by the times marked in the plots. The blue dots are the energy spectrum of the total system consisting of the sensor and the reservoir. The darker-blue dots represent the analytical results from Eq.~\eqref{anaF}, which are in good agreement with the numerical results. The parameters $\omega_{c}=4.5\omega_{0}$, $s=0.5$, $\bar{n}=100$, and $\beta=0.5$ are used.}\label{fig:fig1}
\end{figure*}
\begin{figure*}
\centering
\includegraphics[angle=0,width=.9\textwidth]{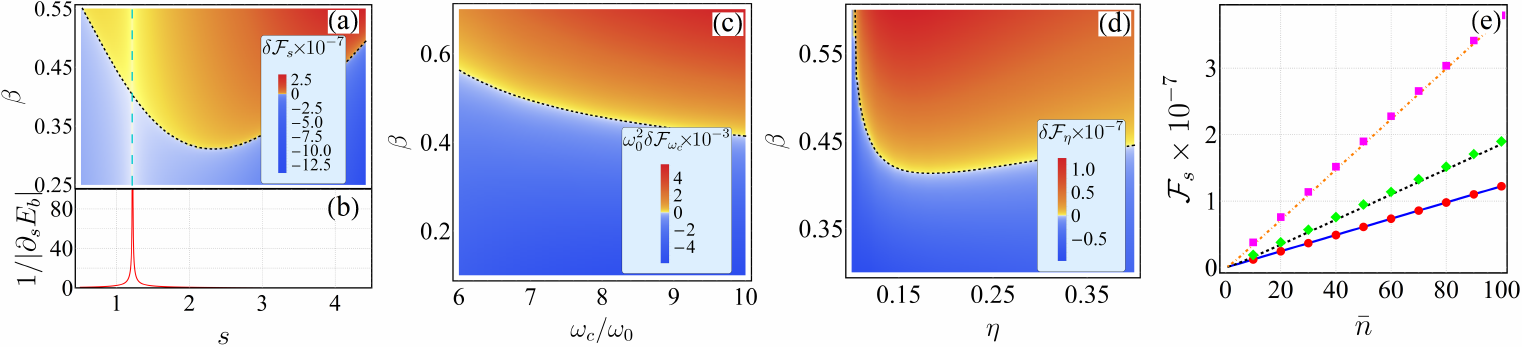}
\caption{(a,c,d) $\delta\mathcal{F}_{\theta}(40\omega_{0}^{-1})$ as a function of $\beta$ and $\theta=s$, $\omega_c$ and $\eta$. The dotted lines are analytically evaluated from $\Theta(\beta,\bar{n})=\bar{n}$. (b) A singular point of the $\mathcal{F}_s$ is present at $s\simeq1.2$. (e) $\mathcal{F}_s(40\omega_{0}^{-1})$ as the change of $n$ when $\beta=0$ (solid line), $0.5$ (dotted line), and $1$ (dot-dashed line). The circles, diamonds, and squares are analytically obtained from Eq. \eqref{anaF}.
The parameters $(\eta,\omega_{c})=(0.4,10\omega_0)$ in panels (a) and (b), $(\eta,s)=(0.2, 1)$ in panel (c), $(s,\omega_c)=(1,10\omega_0)$ in panel (d), and $(\eta,s,\omega_c)=(0.4,3,10\omega_0)$ in panel (e) are used.}\label{fig:fig2}
\end{figure*}
	
\section{Numerical results}
To verify our non-Markovian sensing scheme, we present numerical simulations of the dynamics of QFI. In Fig.~\ref{fig:fig1}, we plot the evolution of the QFI for $\theta=s$, $\omega_c$, and $\eta$ in different $\eta$. We see that, in the regime $\eta\lesssim0.13$ where the bound state is absent, $\mathcal{F}_\theta(t)$ quickly increase to their maxima, and then decrease to zero as time increases. Here the reservoir causes the sensor to undergo a complete decoherence such that its quantum characters are fulled destroyed. This is qualitatively consistent with the Markovian approximate result in Eq. \eqref{MAFI} and the previous works \cite{PhysRevA.97.012126,PhysRevA.97.012125,tamascelli2020quantum,SALARISEHDARAN2019126006,PhysRevA.101.032112}. On the contrary, $\mathcal{F}_\theta(t)$ monotonically increase with time when $\eta\gtrsim0.13$ where the bound state is formed. Well matching the analytical result in Eq. \eqref{anaF} (see the darker-blue dots), $\mathcal{F}_\theta(t)$ in this regime exhibit a square power law with the encoding time. Such law is achievable only in the unitary-encoding quantum metrology in the ideal case \cite{PhysRevLett.79.3865,Gefen2019}. It indicates that although the nonunitary encoding can cause decoherence to the sensor, we still have the chance to obtain a precision scaling with time as ideal as the unitary encoding scheme. Different from many previous studies on sensing a reservoir~\cite{PhysRevA.97.012126,PhysRevA.97.012125,tamascelli2020quantum,SALARISEHDARAN2019126006,PhysRevA.101.032112} in which the precision becomes worse as time increases, our result reveals a mechanism to make the encoding time as a resource in improving the precision. It demonstrates the distinguished role played by the bound state in boosting the QFI of our non-Markovian quantum sensing scheme.

Another role of the bound state is that it permits us to obtain a higher precision than the SNL via optimizing the quantum resource. We define $\delta\mathcal{F}_{\theta}(t)\equiv\mathcal{F}_{\theta}(t)-\mathcal{F}_{\theta}(t)|_{\beta=0}$ as a witness to quantify the effect of squeezing. Focusing on the regime in the presence of the bound state, we obtain $\delta\mathcal{F}_{\theta}(t)=4Z^2t^2(\partial_\theta E_b)^2[\Theta(\bar{n},\beta)-\bar{n}]$ from Eq.~\eqref{anaF}. It can be seen that the precision goes beyond the benchmarked SNL achieved by the coherent state as long as $\Theta(\bar{n},\beta)>\bar{n}$. Figure~\ref{fig:fig2}(a) shows the exact long-time $\delta\mathcal{F}_{s}(t)$ in different $\beta$ and $s$. We really observe a clear threshold perfectly matching the analytical criterion $\Theta(\beta,\bar{n})=\bar{n}$, above which the SNL is surpassed. An exception occurs at $s\simeq1.2$, where $\partial_\theta E_b=0$ [see Fig.~\ref{fig:fig2}(b)]. Such singular point is called Rayleigh's curse~\cite{Gefen2019,PhysRevX.6.031033,PhysRevA.99.013808}, which sets a limit on the optical-imaging resolution. The same result of squeezing-enhanced precision is confirmed by $\delta\mathcal{F}_{\omega_c}$ in Fig.~\ref{fig:fig2}(c) and $\delta\mathcal{F}_\eta$ in Fig.~\ref{fig:fig2}(d). The behavior of $\mathcal{F}_s$ as the change of $\bar{n}$ in Fig.~\ref{fig:fig2}(e) verifies that, although the long-time QFI has the same scaling relation to $\bar{n}$ with the SNL, a sufficient room to boost the QFI by the prefactor still exists. All these numerical results verify our conclusion that the squeezing can enhance the precision to sense the reservoir.

%%%%%%%%%%%%%%%%%%%%%%%%%%%%%%%%%%%%%%%%%%%%%%%%%%%%%
\section{Discussion and conclusions}
%%%%%%%%%%%%%%%%%%%%%%%%%%%%%%%%%%%%%%%%%%%%%%%%%%%%%
Our conclusion in Eq. \eqref{anaF} is independent of the form of spectral density. Although only the Ohmic-family form is considered, our result is generalizable to other cases, where the specific condition of forming a bound state might be different, but the conclusion remains unchanged. The non-Markovian effect has been observed in optical and optomechanical systems \cite{Liu2011,Benardes2015,Gro2015}. The bound state and its effect on the open-system dynamics have been observed in circuit QED \cite{Liu2016} and ultracold atom \cite{Kri2018} systems. These progresses provide a strong support to our scheme. We can verify our conclusion by using the experimental parameters in Ref. \cite{Liu2016}, as shown in the Supplemental Material \cite{SupplementalMaterial}. It indicates that our finding is realizable in the state-of-the-art technique of quantum optics experiments. Our result could be generalized to the finite-temperature reservoir case, where the impact of the bound state on the equilibration dynamics has been found \cite{PhysRevE.90.022122}. Although our result cannot give the optimal measurement observable, the superiority of our bound-state-favored sensing scheme can be demonstrated via a measurement protocol \cite{SupplementalMaterial}. Note that our work has substantial differences from Ref. \cite{PhysRevLett.123.040402}. First, the reservoir here is the target to which we intend to actively measure, while the one in Ref. \cite{PhysRevLett.123.040402} is the source that passively causes unwanted detrimental influence on the ideal unitary-encoding metrology scheme. Second, our goal here is to achieve a high-precision sensing of a reservoir and eliminate the outstanding error-divergence problem, while Ref. \cite{PhysRevLett.123.040402} concentrates on how to retrieve the ideal performance of a quantum metrology scheme from the influence of decoherence.

In summary, we have proposed an optical sensing scheme to estimate the spectral density of a quantum reservoir. A threshold, above which the QFI scales with the encoding time in the same square power law as the noiseless Ramsey-spectroscopy metrology scheme, is found. This is in contrast to the result in the Markovian approximation and the previous works ~\cite{PhysRevA.97.012126,PhysRevA.97.012125,tamascelli2020quantum,SALARISEHDARAN2019126006,PhysRevA.101.032112} that the precision gets deteriorated with time. It is due to constructive interplay between the non-Markovian effect and the sensor-reservoir bound state: The bound state supplies the intrinsic ability and  the non-Markovian effect supplies the dynamical way to achieve the good performance of our nonunitary-encoding sensing scheme. We further find that such interplay can also permit us to achieve a prefactor surpassing the SNL by the squeezing of the sensor. Paving a way to realize a high-precision sensing to the quantum reservoir by the nonunitary-dynamics encoding, our result is helpful in understanding and controlling decoherence caused by the reservoir.

\section{Acknowledgments}
The work is supported by the National Natural Science Foundation (Grants No. 11704025, No. 11875150, and No. 11834005).

\bibliography{reference}

\end{document}

\title{Supplemental Material for ``Non-Markovian Sensing of a Quantum Reservoir''}
\author{Wei Wu}
\affiliation{School of Physical Science and Technology, Lanzhou University, Lanzhou 730000, China}
\author{Si-Yuan Bai}
\affiliation{School of Physical Science and Technology, Lanzhou University, Lanzhou 730000, China}
\author{Jun-Hong An}
\email{anjhong@lzu.edu.cn}
\affiliation{School of Physical Science and Technology, Lanzhou University, Lanzhou 730000, China}

\maketitle

%%%%%%%%%%%%%%%%%%%%%%%%%%%%%%%%%%%%%%%%%%%%%%%%%%%%%
\section{Displacement vector and covariance matrix}
%%%%%%%%%%%%%%%%%%%%%%%%%%%%%%%%%%%%%%%%%%%%%%%%%%%%%
We can calculate the following mean values of operators via solving the exact quantum master equation in the main text or via making the Gaussian integral in the path-integral influence-functional formulism~\cite{PhysRevE.90.022122}
\begin{eqnarray}
\text{Tr}[\hat{a}\varrho(t)]&=&\text{Tr}[\hat{a}^\dag\varrho(t)]^*=u(t)\text{Tr}[\hat{a}\varrho(0)],\\
\text{Tr}[\hat{a}^2\varrho(t)]&=&\text{Tr}[\hat{a}^{\dag2}\varrho(t)]^*=u^2(t)\text{Tr}[\hat{a}^2\varrho(0)],\\
\text{Tr}[\hat{a}^\dag\hat{a}\varrho(t)]&=&|u(t)|^2\text{Tr}[\hat{a}^\dag\hat{a}\varrho(0)].
\end{eqnarray}
For the initial coherent-squeezed state $\varrho(0)=\hat{\mathcal D}(\alpha)\hat{\mathcal S}(r)|0\rangle\langle 0|\hat{\mathcal S}^\dag(r)\hat{\mathcal D}^\dag(\alpha)$, where $\hat{\mathcal D}(\alpha)=\exp(\alpha\hat{a}^{\dagger}-\alpha^{*}\hat{a})$ and $\hat{\mathcal S}(r)=\exp[r(\hat{a}^{2}-\hat{a}^{\dag2})/2]$ with $|0\rangle$ being the vacuum state, it is easy to calculate
\begin{eqnarray}
\text{Tr}[\hat{a}\varrho(0)]&=&\alpha,\\
\text{Tr}[\hat{a}^2\varrho(0)]&=&\alpha^{2}-\sinh r\cosh r,\\
\text{Tr}[\hat{a}^\dag\hat{a}\varrho(0)]&\equiv&\bar{n}=|\alpha|^{2}+\sinh^{2}r.
\end{eqnarray}
According to the definition $d_i(t)=\text{Tr}[\hat{A}_i\varrho(t)]$ and $\sigma _{ij} =\text{Tr}[\varrho \{\Delta \hat{A}_{i},\Delta \hat{A}_{j}^{\dag}\}]$
with $\hat{\mathbf{A}}=(\hat{a},\hat{a}^\dag)^T$ and $\Delta \hat{A}_{i}=\hat{A}_{i}-d_{i}$, we readily obtain
\begin{eqnarray}\label{eq:eqs7}
\pmb{\mathrm{d}}(t)&=&[\alpha u(t),\alpha^{*}u^{*}(t)]^{T},\\
       \pmb{\sigma}(t)&=&\left[
                  \begin{array}{cc}
                    1+2|u(t)|^{2}\sinh^{2}r & -u(t)^{2}\sinh(2r) \\
                    -u^*(t)^{2}\sinh(2r) & 1+2|u(t)|^{2}\sinh^{2}r \\
                  \end{array}
                \right].~~
\end{eqnarray}

\section{Derivation of Eq. (10)}

The encoding dynamics of the optical sensor is uniquely determined by $u(t)$, which satisfies
\begin{equation}
\dot{u}(t)+i\omega_0u(t)+\int_0^\infty f(t-\tau)u(\tau)d\tau=0,\label{ust}
\end{equation}with $u(0)=1$ and $f(t-\tau)=\int_0^\infty J(\omega)e^{-i\omega(t-\tau)}d\omega$. The integro-differential equation \eqref{ust} can be converted into $\tilde{u}(z)=[z+i\omega_0+\int_0^\infty{J(\omega)\over z+i\omega}d\omega]^{-1}$ by a Laplace transform. Then $u(t)$ is analytically solvable via the inverse Laplace transform to $\tilde{u}(s)$, i.e.,
\begin{equation}
u(t)={1\over 2\pi i}\int_{i\sigma+\infty}^{i\sigma-\infty}{e^{-iEt}\over E-\omega_0+\int_0^\infty {J(\omega)\over \omega-E}d\omega}dE,\label{sinverlal}
\end{equation} where $E=iz$ and $\sigma$ is chosen to be larger than all the poles of the integrand~\cite{doi:10.1080/09500349414550381}. It can be done by finding its pole from
\begin{equation}
y(E)=E \label{Ess}
\end{equation}with $y(E)\equiv\omega_0-\int_0^\infty{J(\omega)\over \omega-E}d\omega$. Equation \eqref{Ess} is not else but the one satisfied by the eigenenergies of the total system consisting of the sensor and the quantum reservoir. Thus its solutions give the energy spectrum of the total system. Equation \eqref{Ess} is ill-defined in the regime $E\in[0,+\infty)$. Its singularity points form a branch cut of Eq. \eqref{sinverlal}. Since $y(E)$ is a monotonic decreasing function in the regime $E\in(-\infty,0)$, Eq. \eqref{Ess} has one and only one root $ E_b$ in this regime as long as $y(0)<0$. We call the corresponding eigenstate of this isolated eigenenergy bound state.

\begin{figure}
\centering
\includegraphics[angle=0,width=.55\textwidth]{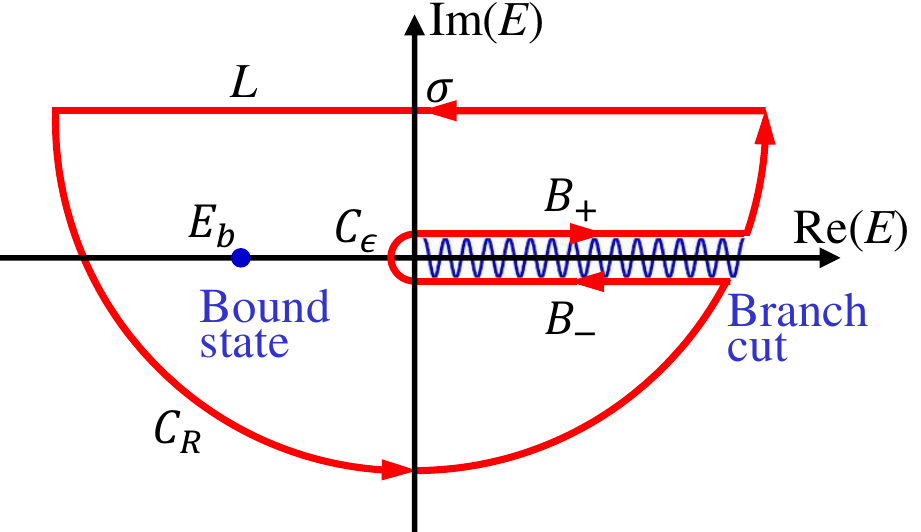}
\caption{ Path of the contour integration in the complex plane $\text{Re}(E)+i\text{Im}(E)$ for the calculation of the inverse Laplace transform of $\tilde{u}(E)$.}\label{contin}
\end{figure}

Figure \ref{contin} shows the path of the contour integration to evaluate Eq. \eqref{sinverlal}. According to the residue theorem, we have
\begin{equation}
\lim_{\epsilon\rightarrow 0,R\rightarrow\infty}\left[\int_{i\sigma+R}^{i\sigma-R}+\int_{C_R}+\int_{B_-}+\int_{C_\epsilon}+\int_{B_+}\right]{e^{-iEt}\over E-\omega_0+\int_0^\infty {J(\omega)\over \omega-E}d\omega}dE=2\pi i\sum_j\text{Res}(E_j),
\end{equation}
where $\text{Res}(E_j)$ represents the residues.  From Jordan's lemma, the integration along the arc paths $C_R$ and $C_\epsilon$ is negligible. Therefore, the paths which have contribution to $u(t)$ are those around the branch cut. Thus we have
\begin{eqnarray}
u(t)&=&\text{Res}(E_b)-{1\over 2\pi i}\lim_{\epsilon\rightarrow0}\left[ \int_{\infty-i\epsilon}^{0-i\epsilon}+\int_{0+i\epsilon}^{\infty+i\epsilon}\right]{e^{-iEt}\over E-\omega_0+\int_0^\infty {J(\omega)\over \omega-E}d\omega}dE\nonumber\\
&=&\text{Res}(E_b)+{1\over 2\pi i}\lim_{\epsilon\rightarrow0}\int_{0}^{\infty}\left[{e^{-iEt}\over E-i\epsilon-\omega_0+\int_0^\infty {J(\omega)\over \omega-E+i\epsilon}d\omega}- {e^{-iEt}\over E+i\epsilon+\omega_0+\int_0^\infty {J(\omega)\over \omega-E-i\epsilon}d\omega}\right]dE,\label{smmmd}
\end{eqnarray}where we have used the fact that there is at most one isolated pole $ E_b$ for the Ohmic-family spectral densities. Equation \eqref{smmmd} can be further simplified by the identities $\lim_{\epsilon\rightarrow0}\int_0^\infty{J(\omega)\over \omega-E\pm i\epsilon}d\omega=-\Delta(E)\mp i\pi J(E)$ with $\Delta(E)=\mathcal{P}\int{J(\omega)d\omega\over E-\omega}$ into
\begin{eqnarray}
u(t)&=&\text{Res}(E_b)+{1\over 2\pi i}\int_{0}^{\infty}\left[{e^{-iEt}\over E-\omega_0-\Delta(E)-i\pi J(E)}- {e^{-iEt}\over E+\omega_0-\Delta(E)+i\pi J(E)}\right]dE\nonumber\\
&=&\text{Res}(E_b)+\int_{0}^{\infty}{J(E)e^{-iEt}\over [E-\omega_0-\Delta(E)]^2+[\pi J(E)]^2}dE.
\end{eqnarray}
For the Ohmic-family spectral densities, the isolated singularity point $E_b$ of the integrand is the first-order pole. Therefore, the residue can be evaluated by the L'Hospital's rule as
\begin{eqnarray}
\text{Res}(E_b)&=&\lim_{E\rightarrow E_b}{(E-E_b)e^{-iEt}\over E-\omega_0+\int_0^\infty {J(\omega)\over \omega-E }d\omega }=\lim_{E\rightarrow E_b}{e^{-iEt}-ite^{-iEt}(E-E_b)\over 1+\int_0^\infty {J(\omega)\over (\omega-E)^2}d\omega}=Z e^{-iE_bt},
\end{eqnarray}where $Z=[1+\int_0^\infty {J(\omega)d\omega\over (\omega-E_b)^2}]^{-1}$. Therefore, we arrive at the final form of $u(t)$ as
\begin{equation}
u(t)=Z e^{-iE_bt}+\int_{0}^{\infty}{J(E)e^{-iEt}\over [E-\omega_0-\Delta(E)]^2+[\pi J(E)]^2}dE.
\end{equation}

\begin{figure}
\centering
\includegraphics[angle=0,width=1.0\textwidth]{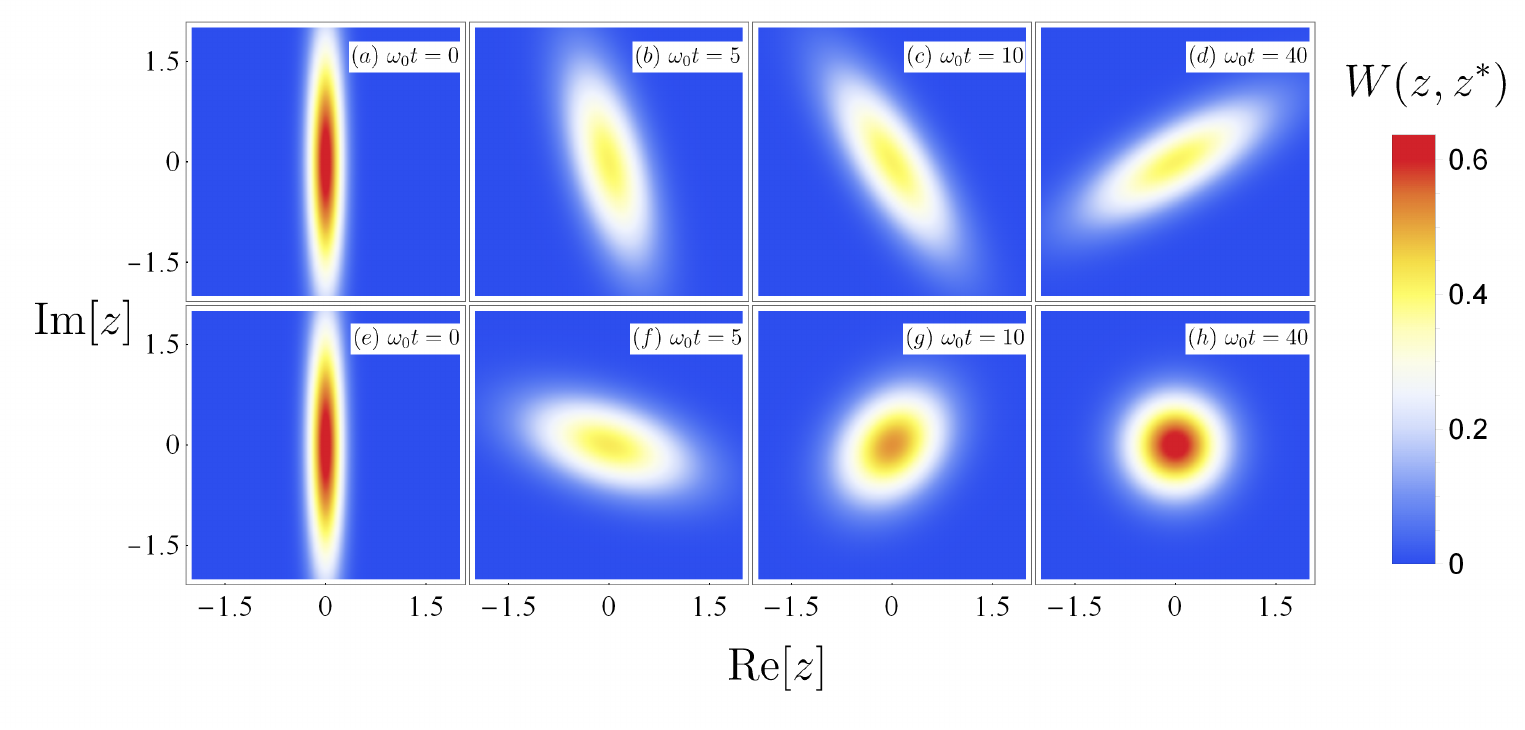}
\caption{ Top panel: Wigner function $W(z,z^{*})$ in the presence of the bound state when $\omega_{c}=20\omega_{0}$ at $t=0$ (a), $5\omega_{0}^{-1}$ (b), $10\omega_{0}^{-1}$ (c) and $40\omega_{0}^{-1}$ (d). Bottom panel: Wigner function $W(z,z^{*})$ in the absence of the bound state when $\omega_{c}=\omega_{0}$ at $t=0$ (e), $5\omega_{0}^{-1}$ (f), $10\omega_{0}^{-1}$ (g) and $40\omega_{0}^{-1}$ (h). Other parameters are $\alpha=0$, $r=1$, $\eta=0.1$ and $s=0.5$. }\label{figwigner}
\end{figure}

\begin{figure}
\centering
\includegraphics[angle=0,width=.8\textwidth]{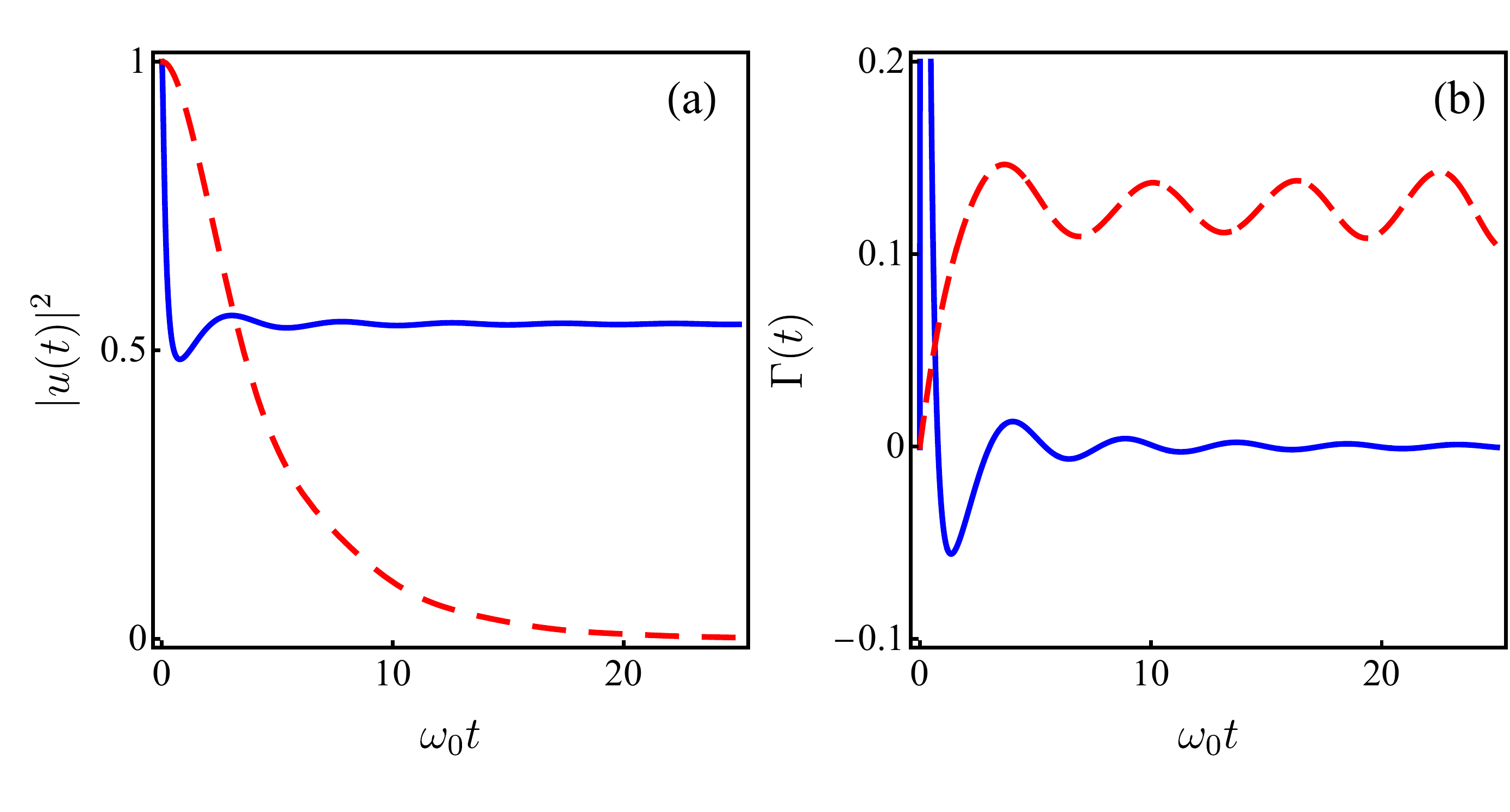}
\caption{(a) Evolution of $|u(t)|^{2}$ with the bound state $\omega_{c}=20\omega_{0}$ (blue solid line) and without bound state $\omega_{c}=\omega_{0}$ (red dashed line). (b) The decay rate $\Gamma(t)$ with the bound state $\omega_{c}=20\omega_{0}$ (blue solid line) and without bound state $\omega_{c}=\omega_{0}$ (red dashed line). The parameters $\eta=0.1$ and $s=0.5$ are used.}\label{derat}
\end{figure}

In order to provide an intuitive picture of the profound influence of the bound state on the dynamics of the sensor, we here plot in Fig.~\ref{figwigner} the time evolution of Wigner function for the initial squeezed state in the presence and absence of the bound state. The Wigner function of the sensor is given by
\begin{equation}
W(z,z^{*})=\frac{1}{\pi\sqrt{|\mathrm{det} \mathbf{C}|}}\exp\bigg{[}-\frac{1}{2}(\mathbf{X}-\mathbf{\bar{X}})^{T}\mathbf{C}^{-1}(\mathbf{X}-\mathbf{\bar{X}})\bigg{]},
\end{equation}
where $\mathbf{X}=(\frac{z+z^{*}}{\sqrt{2}},\frac{z-z^{*}}{\sqrt{2}})^{T}$, $\mathbf{\bar{X}}=(\langle \hat{X}_{1}\rangle,\langle \hat{X}_{2}\rangle)^{T}$ and $\mathbf{C}$ are, respectively, the mean value vector and the covariance matrix with $\mathbf{C}_{ij}=\frac{1}{2}\langle \hat{X}_{i}\hat{X}_{j}+\hat{X}_{j}\hat{X}_{i}\rangle-\langle \hat{X}_{i}\rangle\langle \hat{X}_{j}\rangle$ of the quadrature operators $\hat{X}_{1}=\frac{\hat{a}+\hat{a}^{\dagger}}{\sqrt{2}}$ and $\hat{X}_{2}=\frac{\hat{a}-\hat{a}^{\dagger}}{i\sqrt{2}}$. The explicit forms of $\mathbf{\bar{X}}$ and $\mathbf{C}$ can be calculated as
\begin{eqnarray}
\mathbf{X}&=&\Big{(}\sqrt{2}\mathrm{Re}[\alpha u(t)],\sqrt{2}\mathrm{Im}[\alpha u(t)]\Big{)}^{T},\\
\mathbf{C}&=&\Big{(}\frac{1}{2}+|u(t)|^{2}\sinh^{2}r\Big{)}\mathbf{1}_{2\times 2}-\frac{\sinh(2r)}{2}\left(
                                                                                 \begin{array}{cc}
                                                                                   \mathrm{Re}[u^{2}(t)] & \mathrm{Im}[u^{2}(t)] \\
                                                                                   \mathrm{Im}[u^{2}(t)] & -\mathrm{Re}[u^{2}(t)] \\
                                                                                 \end{array}
                                                                               \right).
\end{eqnarray}
It can be clearly seen from Fig.~\ref{figwigner} that the quantum squeezing can be partially preserved in the steady state if the bound state is formed. On the contrary, quantum squeezing is totally destroyed when the bound state is absent. Two factors play an essential role in this result: The bound state and the non-Markovian dynamics.
First, the bound state supplies the intrinsic ability to suppress the degradation of the quantum resource of the sensor caused by the reservoir. The bound state is a stationary state of the total system including the sensor and the reservoir. Therefore, the quantum squeezing contained in the bound-state component of the initial state is trapped during time evolution. Such result on the dominated role played by the bound state in the nonunitary dynamics of the sensor is confirmed by the Wigner function in Fig.~\ref{figwigner}. Second, the non-Markovian dynamics supplies the dynamical way to suppress the degradation of the quantum resource of the sensor caused by the reservoir. If the Markovian approximation is made, the effect of the bound state would not be manifested in the dynamics. The interplay of the two factors can be equivalently understood by the non-Markovian backflow. The non-Markovian dynamics is governed by the master equation (2) in the main text. The formation of the bound state has two impacts on the time-dependent dissipation rate $\Gamma(t)$ [see Fig. \ref{derat}]. The first impact is that $\Gamma(t)$ can transiently take negative values, which is just a manifestation of the information/energy backflow from the reservoir to the sensor. The second one is that $\Gamma(t)$ asymptotically approaches zero, which ensures the long-time suppression to the reservoir-induced degradation to the quantum coherence of the sensor.


\section{Quantum Fisher information in Markovian limit}

When the sensor-reservoir coupling is weak and the typical time scale of the reservoir correlation function is smaller than the one of the sensor, we can apply the Markovian approximation to the integro-differential equation in the main text and obtain $u_\text{MA}(t)= e^{-\{\kappa+i[\omega_0+\Delta(\omega_{0})]\}t}$ with $\kappa=\pi J(\omega_0)$ and $\Delta(\omega_0)=\mathcal{P}\int _0^\infty {J(\omega)\over \omega_0-\omega}d\omega$. It reduces $\Gamma_\text{MA}(t)= \kappa$ and $\Omega_\text{MA}(t)=\omega_0+\Delta(\omega_0)$~\cite{PhysRevE.90.022122}, which equal to the ones in the Markovian master equation. Then the quantum Fisher information (QFI) can be readily calculated
\begin{equation}
\mathcal{F}_{\theta}^{\text{MA}}(t)=2\bar{n}(\partial_{\theta}\kappa)^{2}t^{2}\Big{\{}\beta[\coth(\kappa t)-1]-\frac{4\beta(\bar{n}\beta+1)}{e^{4\kappa t}+2\bar{n}\beta (e^{2\kappa t}-1)}+\frac{2(1-\beta)}{e^{4\kappa t}+4\bar{n}\beta( e^{2\kappa t}-1)}\Big{[}2\bar{n}\beta +e^{2\kappa t}+2\sqrt{\bar{n}\beta(\bar{n}\beta+1)}\Big{]}\Big{\}}.\label{MAQF}
\end{equation}
Here, we have neglected the constant frequency shift $\Delta(\omega_0)$, which generally can be renormalized into the free frequency $\omega_0$ of the sensor.
One can check $\lim_{t\rightarrow\infty}\mathcal{F}_{\theta}^{\text{MA}}(t)=0$. It means that the message of spectral density of the quantum reservoir vanishes in the long-encoding time limit and no information can be extracted from the steady state of the quantum sensor.

Next, we examine the behavior of Eq. \eqref{MAQF} in the large $\bar{n}$ limit. In this case, one can expand $\mathcal{F}_{\theta}^{\mathrm{MA}}(t)$ in power of $\bar{n}^{-1}$. Its leading term reads
\begin{equation}
\mathcal{F}^\text{MA}_\theta(t)=2(1-\beta)(\partial_\theta\kappa)^2[\coth(\kappa t)-1]\bar{n}t^2.
\end{equation}
Its maximal value with respect to $t$ can be obtained by calculating ${d\mathcal{F}^\text{MA}_\theta(t)\over dt}=0$. It results in
\begin{equation}
2[1-\coth(\kappa t)]+\kappa t\text{csch}(\kappa t)^2=0, \label{largff}
\end{equation}whose solution reads $t=[1+\mathcal{W}(-2/e^{2})/2]\kappa^{-1}=0.80\kappa^{-1}$ with $\mathcal{W}(x)$ being the Lambert $W$-function.
Substituting it into Eq. \eqref{largff}, we obtain the maximal QFI
\begin{eqnarray}
\max_{\beta,t}\mathcal{F}_{\theta}^{\text{MA}}(t)&=&\Big{[}-2 \mathcal{W}(-2/e^{2})-\mathcal{W}^{2}(-2/e^{2})\Big{]}\Big{(}\frac{\partial_{\theta}\kappa}{\kappa}\Big{)}^{2}\bar{n}=0.65(\partial_\theta \ln\kappa)^2\bar{n},\label{MarkF}
\end{eqnarray}
when $\beta=0$.  We see from Eq. \eqref{MarkF} that the maximal $\mathcal{F}_{\theta}^{\mathrm{MA}}$ is proportional to $\bar{n}$. It means that the sensing precision optimized to the encoding time and the quantum resource scales with the total photon number in the same manner as the shot-noise limit in the Markovian approximation. No quantum superiority is delivered by the squeezing and the encoding time in this case.

\section{Exact quantum Fisher information}

As a benchmark to compare with, we first consider the initial state being a coherent state, i.e., $\beta=0$. One can readily derive that the QFI in this case reads $\mathcal{F}_{\theta}(t)|_{\beta=0}=4\bar{n}|\partial_{\theta}u(t)|^{2}$. Focusing on the parameter regime in the presence of the bound state and substituting the long-time form $Ze^{-i E_bt}$ with $Z=[1+\int_0^\infty{J(\omega)d\omega\over( E_b-\omega)^2}]^{-1}$ of $u(t)$ into this formula, we have
\begin{equation}
\lim_{t\rightarrow\infty}\mathcal{F}_{\theta}(t)|_{\beta=0}=4\bar{n}|\partial_{\theta}Z-iZ(\partial_{\theta} E_{b})t|^2\simeq4 Z^{2}\bar{n}(\partial_{\theta} E_{b})^{2}t^{2},\label{Cohetin}
\end{equation}where the time-independent term $\partial_\theta Z$ has been neglected in the long-time condition.

In the general case with $\beta\neq 0$, some straightforward algebra can lead to
\begin{eqnarray}
\lim_{t\rightarrow\infty}\mathcal{F}_{\theta}(t)&\simeq&4Z^{2}(\partial_{\theta} E_{b})^{2}\Theta(\bar{n},\beta)t^{2},\label{squin}\\
\Theta(\bar{n},\beta)&=&\frac{2Z^{2}\bar{n}\beta(1+\bar{n}\beta)}{1+2\bar{n}\beta Z^{2}(1-Z^{2})}+\frac{\bar{n}(1-\beta)[1-2Z^{2}(\sqrt{\bar{n}\beta(1+\bar{n}\beta)}-\bar{n}\beta)]}{1+4\bar{n}\beta Z^{2}(1-Z^{2})}.
\end{eqnarray}
It is easy to check that $\Theta(\bar{n},\beta)$ reduces to $\bar{n}$ and Eq. \eqref{squin} returns to Eq. \eqref{Cohetin} when $\beta=0$. In the large $\beta\bar{n}$ limit, we can expand $\Theta(\bar{n},\beta)$ in power of $(\beta\bar{n})^{-1}$ and keep the leading term as $\Theta(\bar{n},\beta)\simeq\frac{\bar{n}\beta}{1-Z^{2}}$.

In the following, we give the explicit form of $\partial_\theta E_b$ for $\theta=\eta$, $\omega_c$, and $s$. Using the fact that $ E_b<0$ for the Ohmic-family spectral density $J(\omega)=\eta\omega(\omega/\omega_c)^{s-1}e^{-\omega/\omega_c}$, we have
\begin{equation*}
\int_0^\infty{J(\omega)\over \omega- E_b}d\omega=\eta\omega_{c}\mathrm{Ei}_{s+1}(- E_b/\omega_c)\underline{\Gamma}(s+1)e^{- E_b/\omega_c},
\end{equation*}
where $\mathrm{Ei}_{n}(z)\equiv\int_{1}^{\infty}e^{-zt}t^{-n}dt$ is the exponential integral function and $\underline{\Gamma}(s)$ is the Euler's gamma function. Making the derivative of $\theta=\eta$, $\omega_c$, and $s$ to the equation satisfied by $ E_b$, i.e., $\omega_{0}- E_{b}-\int_0^\infty{J(\omega)\over \omega- E_b}d\omega=0$, we can calculate
\begin{eqnarray}
\partial_{\eta} E_{b}&=&\frac{\omega_{c}\mathrm{Ei}_{s+1}(- E_{b}/\omega_c)\underline{\Gamma}(s+1)}{\eta \mathrm{Ei}_{s+1}(- E_{b}/\omega_c)\underline{\Gamma}(s+1)-\eta\mathrm{Ei}_{s}(- E_{b}/\omega_c)\underline{\Gamma}(s)-e^{ E_{b}/\omega_c}},\\
\partial_{\omega_{c}} E_{b}&=&\frac{\eta\underline{\Gamma}(s)[\omega_{c}(\omega_{c}+ E_{b})e^{ E_{b}/\omega_c}
+ E_{b}(\omega_{c}+ E_{b}-s\omega_{c})\mathrm{Ei}_{s}(- E_{b}/\omega_c)]}{\eta\omega_{c}( E_{b}-s\omega_{c})\underline{\Gamma}(s)\mathrm{Ei}_{s}(- E_{b}/\omega_c)
+\omega_{c}^{2}[\eta\underline{\Gamma}(s)-1]e^{ E_{b}/\omega_c}},\\
\partial_{s} E_{b}&=&\frac{\eta\omega_{c}\underline{\Gamma}(s+1)\big{[}\mathrm{G}_{2,3}^{3,0}\big{(}- E_{b}/\omega_c\Big{|}\begin{array}{c}
                                                                                                                s+1,s+1 \\
                                                                                                                0,s,s \\
                                                                                                              \end{array}                                                                                                            \big{)}-\psi(s+1)\mathrm{Ei}_{s+1}(- E_{b}/\omega_c)\big{]}}{e^{ E_{b}/\omega_c}+\eta\underline{\Gamma}(s+1)[ \mathrm{Ei}_{s}(- E_{b}/\omega_c)-\mathrm{Ei}_{s+1}(- E_{b}/\omega_c)]},
\end{eqnarray}
where
\begin{equation}
\begin{split}
\mathrm{G}_{m,n}^{p,q}\Big{(}z\Big{|}\begin{array}{c}
                                                                                                                a_{1},...,a_{p} \\
                                                                                                                b_{1},...,b_{q} \\
                                                                                                              \end{array}                                                                                                            \Big{)}\equiv\frac{1}{2\pi i}\int_{\mathcal{C}}\frac{\underline{\Gamma}(1-a_{1}-\theta)...\underline{\Gamma}(1-a_{n}-\theta)\underline{\Gamma}(b_{1}+\theta)...\underline{\Gamma}(b_{m}+\theta)}{\underline{\Gamma}(a_{n+1}+\theta)...\underline{\Gamma}(a_{p}+\theta)\underline{\Gamma}(1-b_{m+1}-\theta)...\underline{\Gamma}(1-b_{q}-\theta)}z^{-\theta}d\theta
\end{split}
\end{equation}
is the Meijer G-function ~\cite{Gradshteyn} and $\psi(z)\equiv\partial_{z}[\ln\underline{\Gamma}(z)]$.

During the numerical calculation to the exact QFI, the derivative of $u(t)$ to $\theta$ is performed by the fourth-order finite difference method
\begin{equation}
\partial_{\theta}u(t)\simeq\frac{-u(t)|_{\theta+2\epsilon}+8u(t)|_{\theta+\epsilon}-8u(t)|_{\theta-\epsilon}+u(t)|_{\theta-2\epsilon}}{12\epsilon}
\end{equation}
with $\epsilon=10^{-7}$.

\section{A Specific Measurement Scheme}

\begin{figure}
\centering
\includegraphics[angle=0,width=0.65\textwidth]{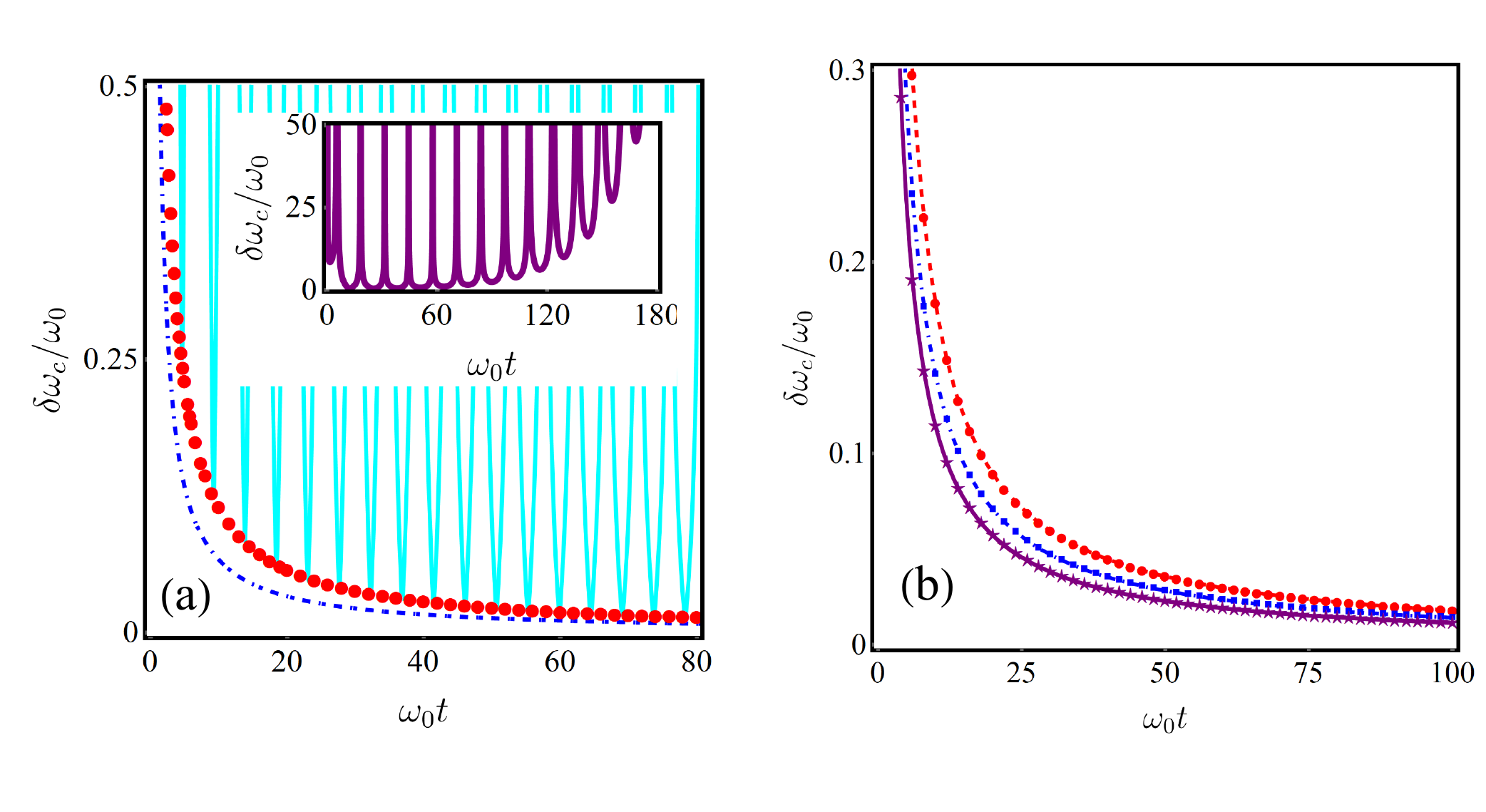}
\caption{(a) Sensing precision with respect to the measurement operator $\hat{\mathcal{O}}=\hat{a}^{\dagger}+\hat{a}$ when $\eta=2$ and $r=2.5$ (the cyan solid line). The red circles are the analytical solution of the local minima obtained from Eq.~\eqref{error}. The blue dashed line is the ultimate precision obtained from the QFI. The inset shows the precision with respect to $\mathcal{\hat{O}}$ in the absence of bound state when $\eta=0.07$. (b) Local minima of the sensing precision when $r=0$ (red dashed line), $0.5$ (blue dotdashed line), and $2.5$ (purple solid line). The circles, squares, and stars are the corresponding analytical results obtained from Eq.~\eqref{error}. Other parameters are chosen as $|\alpha|=2.5$, $\phi=\pi/2$, $\omega_{c}=10\omega_{0}$, and $s=1$.}\label{figm}
\end{figure}

We have studied in the main text the QFI, which refers to the ultimate precision with respect to the optimal measurement observable. The QFI itself cannot give the optimal measurement operator. Therefore, it is necessary to discuss a specific measurement scheme from experimental perspective. The sensing precision with respect to a given measurement operator $\mathcal{\hat{O}}$ can be obtained via the error propagation formula
\begin{equation}
\delta\theta(\mathcal{\hat{O}})=\frac{\delta\mathcal{O}}{|\partial_{\theta}\mathcal{\bar{O}}|},
\end{equation}
where $\delta\mathcal{O}\equiv(\overline{\mathcal{O}^{2}}-\bar{\mathcal{O}}^{2})^{1/2}$ with $\bar{\mathcal{O}}\equiv \mathrm{Tr}[\varrho(t)\mathcal{\hat{O}}]$.

As an illustrative example, we assume $\mathcal{\hat{O}}=\hat{a}^{\dagger}+\hat{a}$. The corresponding sensing precision is given by
\begin{equation}
\delta\theta(\mathcal{\hat{O}})=\frac{\sqrt{2|u(t)|^{2}\sinh^{2}(r)+1-\sinh(2r)\mathrm{Re}[u^{2}(t)]}}{|\alpha[\partial_{\theta}u(t)]+\alpha^{*}[\partial_{\theta}u^{*}(t)]|},
\end{equation}
where $\alpha=|\alpha|e^{i\phi}$. In the Markovian limit, $u(t)\simeq u_\text{MA}(t)= e^{-(\kappa+i\omega_0)t}$, one can easily check that $\delta\theta(\mathcal{\hat{O}})$ becomes larger and larger as time increases. This result means the sensing scheme totally breaks down in the long-encoding time regime, which is the outstanding error-divergency problem. In the presence of the bound state, we have the asymptotic solution $u(t)\simeq Ze^{-iE_{b}t}$ and
\begin{equation}
\delta\theta(\mathcal{\hat{O}})\simeq\frac{\sqrt{1+2Z^{2}\sinh^{2}(r)-Z^{2}\sinh(2r)\cos(2E_{b}t)}}{2Z|\alpha(\partial_{\theta}E_{b})\sin(E_{b}t-\phi)|t}.
\end{equation}
Choosing $\phi=\pi/2$ and $E_{b}t=n\pi$, we can obtain the optimal sensing precision by measuring $\mathcal{\hat{O}}=\hat{a}^{\dagger}+\hat{a}$ as
\begin{equation}~\label{error}
\min\delta\theta(\mathcal{\hat{O}})=\frac{\sqrt{1+2Z^{2}\sinh^{2}(r)-Z^{2}\sinh(2r)}}{2Z|\alpha(\partial_{\theta}E_{b})|t}.
\end{equation}
Its scaling behavior with the encoding time is the same as the ultimate accuracy obtained by QFI in the main text. In Fig.~\ref{figm}(a), we numerically calculate the exact sensing precision $\delta\omega_{c}$ with respect to $\mathcal{\hat{O}}=\hat{a}^{\dagger}+\hat{a}$. One can see that $\delta\omega_{c}$ oscillates with the encoding time and has multiple local minima, which perfectly match with our analytical result in Eq.~\eqref{error}. In the absence of the bound state, these minima gradually become divergent as $\omega_{0}t$ approaches to infinity. However, as long as the bound state is formed, the local minima decrease with the encoding time. The numerical result is in good agreement with our analytical solution \eqref{error}. We also plot in Fig.~\ref{figm}(b) the influence of the initial quantum squeezing on the sensing precision. It is found that the sensing performance can be boosted by increasing the squeezing strength. Although the sensing precision via measuring $\mathcal{\hat{O}}=\hat{a}^{\dagger}+\hat{a}$ does not saturate the Cram\'{e}r-Rao bound governed by QFI [see the blue dotdashed line in Fig.~\ref{figm} (a)], it sufficiently demonstrates the superiority of our bound-state-based sensing scheme in an experimentally friendly manner.

\section{Physical Realization}

\begin{figure}
\centering
\includegraphics[angle=0,width=0.65\textwidth]{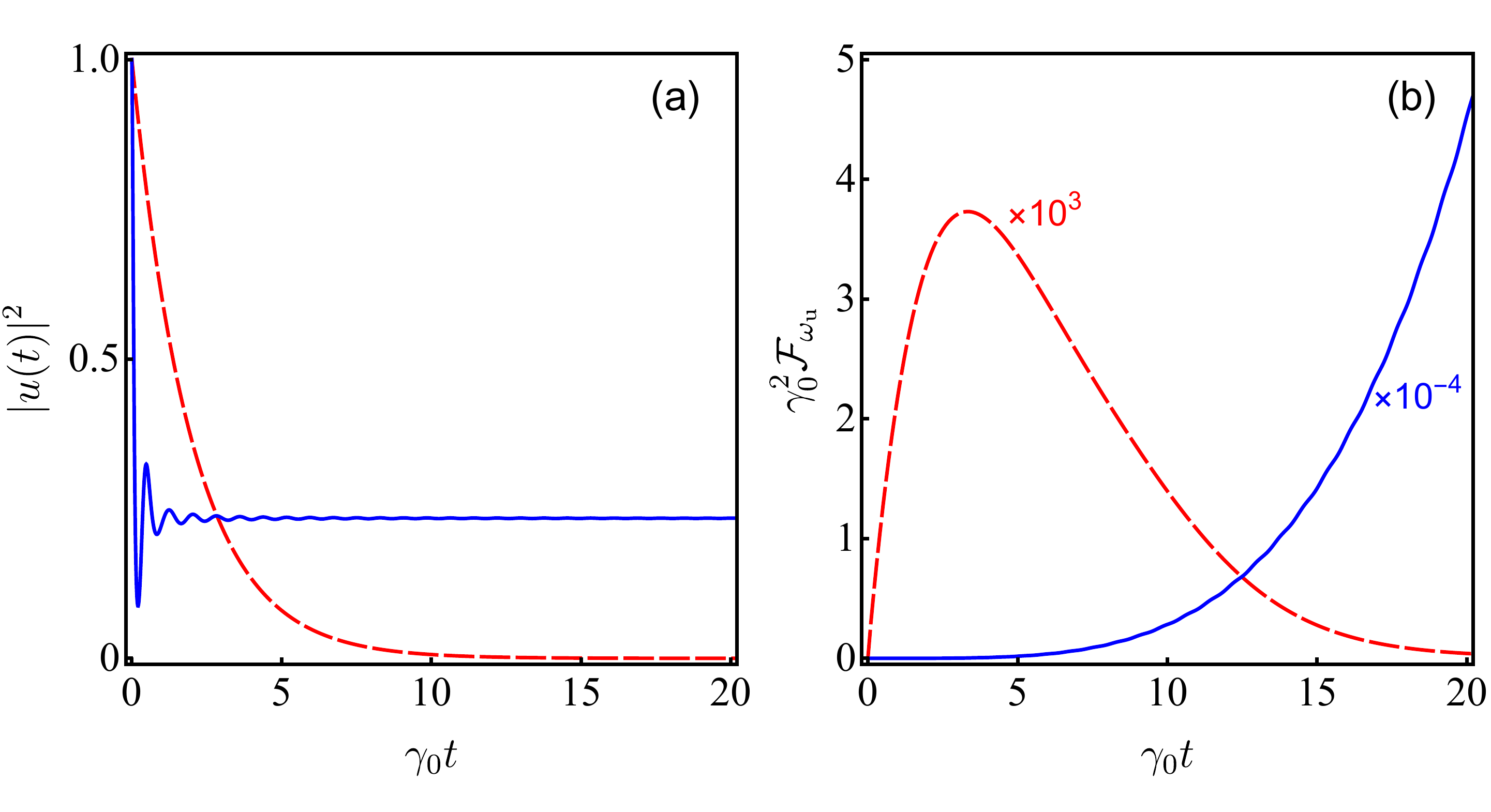}
\caption{Evolution of $|u(t)|^{2}$ in (a) and $\mathcal{F}_{\omega_{\mathrm{u}}}$ in (b) when $\omega_{0}=159\gamma_{0}$ (blue solid line) and $600\gamma_{0}$ (red dashed line). Other parameters are $\omega_{\mathrm{u}}=160\gamma_{0}$, $\bar{n}=100$, and $\beta=0.5$.}\label{figpc}
\end{figure}

A promising physical system to test our theoretical prediction is photonic crystals, which has been recently realized experimentally in a circuit quantum electrodynamics architecture. The dispersion relation of the band-gapped reservoir in the photonic crystal setting is given by $\omega_{k}=\omega_{\mathrm{u}}+\mathbb{A}(k-k_{0})^{2}$, where $\omega_{\mathrm{u}}$ is the band edge frequency and $\mathbb{A}=\omega_{\mathrm{u}}/k_{0}^{2}$ with $k_{0}\simeq\omega_{\mathrm{u}}/c$. In the photonic-crystal-reservoir case, the expression of $u(t)$ can be exactly derived~\cite{doi:10.1080/09500349414550381,Wang_2017}
\begin{equation}
u(t)=e^{-i\omega_{0}t}\sum_{\ell=1}^{3}p_{\ell}e^{(i\delta+\varepsilon x_{\ell}^{2})t}\Big{[}x_{\ell}+\sqrt{x_{\ell}^{2}}\mathrm{Erf}\Big{(}\sqrt{\varepsilon x_{\ell}^{2}t}\Big{)}\Big{]},
\end{equation}
where $\mathrm{Erf}(\cdot)$ is the error function, $\varepsilon=\omega_{\mathrm{u}}(\pi\gamma_{0}/2\omega_{0})^{2/3}$ with $\gamma_{0}$ being the vacuum spontaneous emission rate, $\delta=\omega_{0}-\omega_{\mathrm{u}}$ and $p_{\ell}=x_{\ell}/[(x_{\ell}-x_{\imath})(x_{\ell}-x_{\jmath})]$ with $\imath\neq\jmath\neq\ell=1,2,3$ being the roots of the characteristic equation $(\varepsilon x^{2}+i\delta)\sqrt{\varepsilon}x-(i\varepsilon)^{3/2}=0$. The condition of forming a bound state in the sensor-photonic-crystal reservoir is $\delta<0$~\cite{doi:10.1080/09500349414550381,Wang_2017}. According to recent circuit QED experiments~\cite{PhysRevA.86.023837,Houck2012,Liu2016}, the band edge frequency $\omega_{\mathrm{u}}\simeq 8.0~\text{GHz}$, the vacuum spontaneous emission rate $\gamma_{0}\simeq 50~\mathrm{MHz}$, and the frequency of a single-mode cavity $\omega_{0}\simeq 6.0\mathrm{GHz}$ to $8.5\mathrm{GHz}$. From these parameters, one can immediately find that the condition of forming the bound state can be satisfied. As the illustrative example, we use our sensing scheme to estimate the band edge frequency of the photonic crystal reservoir. In Fig.~\ref{figpc}, we plot the dynamics of $|u(t)|^{2}$ and the corresponding QFI with experimentally achievable parameters. In the presence of the bound state, $|u(t)|^{2}$ approaches a non-zero steady value and the corresponding QFI increases with time in power law. On the contrary, both $|u(t)|^{2}$ and the QFI vanish in the long-encoding-time limit when the bound state is absent. This result is consistent with our conclusion in the main text and demonstrates our bound-state-based sensing scheme is realizable in current experimental condition.

\bibliography{reference}